\documentclass[twocolumn]{aastex631}

\newcommand{\mjup}{M$_{J}$}
\usepackage{float}
\usepackage{appendix}
      
\begin{document}   

\title{Next-Generation Improvements in Giant Exoplanet Evolutionary and Structural Models}

\shorttitle{Next-Generation Improvements in Exoplanet Modeling}
\shortauthors{Sur et al.}

\correspondingauthor{Ankan Sur, Roberto Tejada Arevalo}
\email{ankan.sur@princeton.edu, arevalo@princeton.edu}

\author[0000-0001-6635-5080]{Ankan Sur}
\affiliation{Department of Astrophysical Sciences, Princeton University, 4 Ivy Lane,
Princeton, NJ 08544, USA}
\affiliation{Department of Earth, Planetary, and Space Sciences, University of California Los Angeles, 595 Charles E Young Dr E, LA, CA 90095}
\author[0000-0001-6708-3427]{Roberto Tejada Arevalo}
\author[0000-0002-3099-5024]{Adam Burrows}
\author[0000-0003-3792-2888]{Yi-Xian Chen}
\affiliation{Department of Astrophysical Sciences, Princeton University, 4 Ivy Lane,
Princeton, NJ 08544, USA}


\begin{abstract}

\textcolor{black}{Many evolutionary models of giant exoplanets still rely on simplifying assumptions that are no longer adequate given detailed constraints from Jupiter, Saturn, and modern exoplanet observations. Here, we identify the key physical improvements required for next-generation planetary evolution models using our code, \texttt{APPLE}, which enables systematic emulation and extension of legacy studies. We quantify the effects of updated equations of state, helium rain, fuzzy cores, non-adiabatic and compositionally inhomogeneous envelopes, and improved atmospheric boundary conditions by first isolating the impact of each physical ingredient and then constructing combined baseline models for planets with masses between 0.3 and 4~$M_{\rm Jup}$ to assess their collective influence on planetary structure and observable properties.} We find that the adoption of modern equations of state and realistic heavy-element distributions leads to systematic, but sometimes subtle, differences ($\sim 5$ to 10\%) in radius evolution, while helium rain and the treatment of convection can significantly alter thermal histories and atmospheric compositions (by $\sim$ 5 to 20\%). These updated physical processes must be incorporated into the next-generation exoplanet evolutionary models to achieve physically consistent interpretations of planetary observations.
\end{abstract}
    

\section{Introduction}
 
Structural and evolutionary models of the solar system gas giants Jupiter and Saturn (with masses of 318 and 95 earth masses, respectively) have a long pedigree. Given their masses, it has long been understood that their large radii required hydrogen to be their primary constituent \citep{Demarcus1958, Peebles1964, Zapolsky1969, Hubbard1969, Bodenheimer1976}. Their small inferred thermal conductivities \citep{Hubbard1966, Hubbard1968, Hubbard1973, StevensonSalpeter1977a} suggested that convection dominates their energy transport and cooling from birth \citep{Hubbard1968, Hubbard1970, Grossman1972, Graboske1975}. Their formation out of the protostellar/protoplanetary disk by the assembly of a critical mass heavy-element core (of perhaps $\sim$10 earth masses) that nucleated the subsequent rapid accumulation of hydrogen and helium gas to achieve a gas giant had been (and remains, in broad outline) the paradigm for their formation \citep{Mizuno1980, Stevenson1982, Bodenheimer1986, Lissauer1987, Pollack1996}. Hence, what emerged was a model with a fractionally small heavy-element core surrounded by a convective/adiabatic envelope of predominantly hydrogen and helium (in roughly solar ratios), with a small admixture of heavier elements (in perhaps solar ratios). Convection would homogenize the envelope and make it isentropic. The thin outer atmosphere would act as the valve for energy loss.

This model of a core (that could be thermally inert) with an extensive gas envelope, capped by a molecular atmosphere, has been the conceptual framework for gas giants for decades, epitomized by the pioneering early structural and evolutionary work of Hubbard \citep{Hubbard1968, Hubbard1969, Hubbard1970, Hubbard1977}. After formation, such a model is evolved with an ordinary differential equation for the time derivative of the envelope entropy, an equation of state, and an atmosphere boundary condition \citep{Burrows1993, Burrows1995, Guillot1995, Burrows1997, Baraffe1998, Baraffe2003, Fortney2003, Fortney2004, Aras2006, Fortney2007, Fortney2011, Saumon2008, Phillips2020, Marley2021}. An approximation for the thermal effect of solar insolation is attached (whether incorporated into the atmosphere calculation or in an ad hoc fashion), and the time dependence of the planet's radius, entropy, and luminosity is calculated.

However, recent formation models have suggested the heavy-element core could be extended \citep{Lozovsky2017, helledstevenson2017, Stevenson2022, Bodenheimer2025} and the gaseous hydrogen/helium envelope could grow coevally with the accumulation of solid planetesimals and/or pebbles.  Moreover, recent Juno \citep{Bolton2017a} and Cassini \citep{Spilker2019, Iess2019} probe data strongly suggest there are stably-stratified/non-convective regions in both planets and that their envelopes are not homogeneous. They seem to have ``fuzzy cores" with extended non-flat distributions of heavy elements \citep{Wahl2017, Debras2019, Militzer2022, Mankovich2021, Howard2023b, Militzer2024}. Given this, the traditional structural and evolutionary models are inadequate, and techniques akin to those employed in stellar evolution are required to model them temporally. This has recently been attempted by our group \citep{Tejada2025, Sur2024a, Sur2025a, Sur2025b} and by others \citep{Vazan2018, Muller2020b, Stevenson2022}. 

In light of these developments, a reassessment of literature models for giant exoplanet evolution that adopted the traditional approach employed for solar-system giants would seem in order. Those evolutionary models \citep{Burrows1995, Burrows1997, Burrows2001, Baraffe2003, Fortney2007, Saumon2008, Phillips2020, Marley2021} have underpinned the global interpretation of observations of giant exoplanets over three decades. However, they generally incorporate the thermally inert core plus fully convective envelope paradigm and use ordinary differential equation solvers for the quasi-hydrostatic evolution of the envelope's entropy and luminosity. \textcolor{black}{These evolutionary models have also neglected semiconvective heat transport. \citet{Leconte2012} showed that introducing compositional gradients and allowing departures from strictly adiabatic interiors can significantly affect planetary cooling histories, even though heat and compositional transport were not treated self-consistently in their formalism. Due to the lack of an adequate implementation of semiconvection in current stellar and planetary evolution codes, we neglect its effects in this study and leave a detailed treatment for future work.} In addition, these exoplanet models do not incorporate helium rain \citep{Burrows1997, Baraffe2003, Aras2006, Fortney2007,Saumon2008, Phillips2020, Marley2021} and have generally employed the older equation of state (EOS) \citep{Saumon1995} for hydrogen/helium mixtures. Finally, they frequently ignored the contribution of the heavy elements to the envelope EOS, ignoring the contribution of the heavies to the specific heats and the pressures \textcolor{black}{(with the exception of \citet{Baraffe2008}).} 

With this paper, we set out to explore the differences between heritage giant exoplanet models and those that incorporate updated equations of state, fuzzy core and stably-stratified heavy-element envelopes, helium rain, and updated atmospheric boundary conditions. We here do not set out to provide an extensive grid of such models for general community use. This will follow later.  We merely highlight the differences between published models and representative structural and evolutionary models, channeling the more modern perspective (however provisional). To accomplish this, we use the \texttt{APPLE} code \citep{Sur2024a}.  We will see that there can be quantitative differences in the predicted observables, particularly for the lower mass giant exoplanets.

In \S\ref{sec:methods}, we describe our choice of hydrogen and helium (H-He) equation of state (EOS), heavy element equation of state, atmosphere boundary conditions, and our handling of heat and composition transport. We reproduce past models with their assumptions and EOSes in \S\ref{sec:reproduction}. We present improved best practices for modeling the evolution of gas giant planets in \S\ref{sec:improvements}, and compute a sample of new evolution models in \S\ref{sec:new_models}. We provide concluding remarks in \S\ref{sec:conclusion}. 

\section{Methods}\label{sec:methods}

In the study of giant planets, the H-He EOS plays a central role in the evolution of their effective temperatures, radii, and interior temperature profiles \citep{Militzer2013, Chabrier2019}. The H-He EOS of \citet[][SCvH95]{Saumon1995} has been widely used in gas giant planet evolution models, but since such early work did not account for non-ideal entropy and volume interaction terms in the volume-addition law prescriptions \citep{Saumon1995}, it has now been superseded by H-He EOSes that do include such terms \citep[See][and discussions therein]{Tejada2024}. The H-He EOS of \citet[][CMS19]{Chabrier2019} incorporates these effects in the EOS published by \citet[][CD21]{Chabrier2021}. The CD21 H-He EOS captures the non-ideal mixing effects of the H-He EOS calculated by \cite{Militzer2013}. Most recently, \citep{Howard2023a} included these non-ideal effects as a function of helium fraction.  
The differences between the CD21 EOS and the updates to the CMS19 by \cite{Howard2023a} are minor, as shown in Figure 4 of \cite{Howard2023a} and Figure 1 of \cite{Tejada2024}. We further note here that recent experimental results by \citet{Liu2025} show good agreement with both the CD21 EOS and the \cite{Howard2023a} results. We therefore use the updated H-He EOS of CD21 in later sections to remain consistent with our previous work \citep[e.g.,][]{Tejada2025, Sur2025a}.

The metal ($Z$) EOS is here represented by the multi-phase water calculations of \citet[][AQUA]{Haldemann2020}, which includes the calculations of \cite{Mazevet2019}.\footnote{As \cite{Aguichine2025} pointed out, the water EOS of \cite{Mazevet2019} contained an error in the non-ideal entropy terms which did not affect the density. This has then been corrected \citep{Mazevet2021}, but no other comprehensive water EOS tables are currently available. Our own calculations and those of \cite{Tejada2025b} suggest that these corrections are small and inconsequential to any conclusions in this work.} Throughout this work, the solar metallicity (1 $Z_\odot$) is taken to be 0.0156 by mass. We distinguish between \textit{metallicity} and \textit{metal fraction} where metallicity is quoted as a ratio to solar values, and metal fractions are always metal mass fractions.\footnote{For example, 10, 100, and 1000 $Z_\odot$ metallicities are 0.136, 0.613, and 0.941 metal fraction by mass, respectively, assuming 1 $Z_\odot = 0.0156$. }

Heat and composition transport is primarily treated via mixing-length-theory convection under the Ledoux criterion \citep{Ledoux1947}.
Any region which meets the Ledoux criterion is convective, and regions which do not meet this criterion are stable to convection. In stable regions, conduction and microscopic diffusion are assumed to be the dominant modes of heat and composition transport. Our evolution code \texttt{APPLE} and its heat and composition transport methodology is fully described in \cite{Sur2024a}. Moreover, a comprehensive discussion on our applications of the Ledoux criterion is found in Section 2 of \cite{Tejada2025}. In \S\ref{sec:new_models}, we discuss the implications of using the Schwarzschild criterion \citep{schwarzschild1965}.


The effects of semiconvection, or ``double-diffusive'' convection, are omitted in this study due to the existing uncertainties in its treatment and in the relevant microphysical properties. While semiconvection could affect the evolution of helium rain layers and stably-stratified regions in gas giant fuzzy cores, their impact on 1-D evolution models may be minor \citep[See Figure 8 of][]{Muller2020b}, though this is still an open question.

\begin{figure*}
    \centering
    \includegraphics[width=0.92\linewidth]{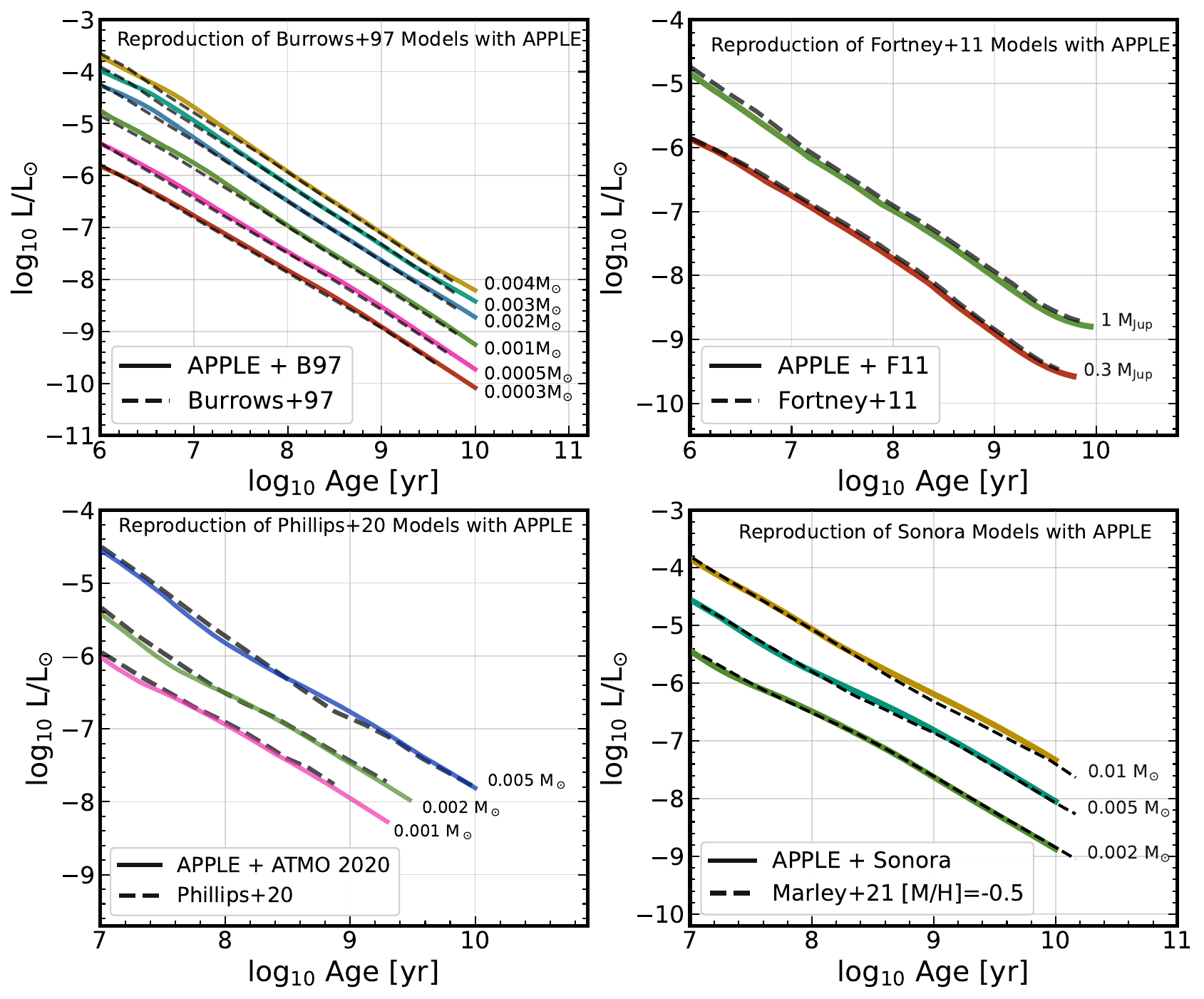}
    \caption{\textcolor{black}{Reproduction of various heritage exoplanet evolutionary models (dashed black curves) using the model setups and inputs in those papers, as computed with \texttt{APPLE} (solid colored lines) for a range of planetary masses. Masses are indicated next to each line (in units of solar masses) and luminosities are in solar units. Unless explicitly stated otherwise, all models are adiabatic, have no solid core, neglect rotation, and neglect helium rain. Top left: Evolutionary sequences from \citet{Burrows1997} that uses the isolated, non-gray model atmosphere grids. Top right: Evolutionary sequences computed using the irradiated atmospheric boundary tables of \citet{Fortney2011} for Jupiter- and Saturn-mass planets, assuming 3.16 $Z_{\odot}$ metallicity. Bottom left: Models from \citet{Phillips2020}, which extend the earlier \citet{Baraffe2003} evolutionary calculations. In \texttt{APPLE}, we adopt the ATMO atmospheric grids of \citet{Phillips2020} with the CMS19 H–He EOS, using an effective helium fraction $Y^{\prime} = Y + Z$ to approximate heavy elements, as in their work. Bottom right: Sonora-Bobcat models \citep{Marley2021} at [M/H]$=-0.5$ are reproduced using their published atmospheric boundary conditions and interior assumptions.}}
    \label{fig:apple-reproduction}
\end{figure*}

\section{Reproduction of Past Evolutionary Models}
\label{sec:reproduction}

Most evolutionary calculations in the past assumed adiabatic interiors, omitted solid cores, neglected rotation, and did not account for the effects of helium rain. Here, we reproduce a series of exoplanet evolutionary histories from various groups using the planetary evolution code \texttt{APPLE} \citep{Sur2024a}, adopting their same assumptions, for a range of planetary masses spanning from Saturn mass up to a few Jupiter masses as shown by the colored lines in Figure~\ref{fig:apple-reproduction}. This exercise is conducted to ensure we can emulate past models and practices. The top-left panel presents the evolutionary sequences from \citet{Burrows1997}, where the envelope helium mass fraction is $Y = 0.25$ and the interior is modeled using the SCvH95 equation of state \citep{Saumon1995}. Atmospheric boundary conditions are provided by the isolated, non-gray model atmosphere grids of \citet{Burrows1997}. Their study modeled solar-metallicity objects spanning 0.3 M$_{\rm Jup}$ to $70\, M_{\rm Jup}$ (down to $T_\mathrm{eff} \sim 100$~K) using a sophisticated atmospheric boundary condition that employed a detailed molecular opacity and equilibrium database. The top-right panel displays the reproduction of evolutionary sequences based on the irradiated atmospheric boundary tables of \citet{Fortney2011} for Jupiter- and Saturn-mass planets, assuming a metallicity of 3.16 times the solar value and adopting the SCvH95 EOS for the hydrogen–helium mixture. In these models, the Jupiter case includes a $10 \,M_{\oplus}$ core composed of iron and post-perovskite, with an envelope heavy-element mass fraction $Z = 0.059$ and employing the old AQUA EOS, while the Saturn case assumes a $21 M_{\oplus}$ core with a $Z = 0.03$ envelope. The bottom-left panel presents the reproduction of models from \citet{Phillips2020}, which extend the earlier work of \citet{Baraffe2003}. In our implementation within \texttt{APPLE}, the ATMO atmospheric grids from \citet{Phillips2020} are combined with the CMS19 H–He EOS \citep{Chabrier2019} for the envelope. Rather than using a dedicated heavy-element EOS, these models mimic the effect of heavy elements by adopting an effective helium fraction, $Y^{\prime} = Y + Z = 0.2919$, which increases the mean molecular weight of the mixture. Finally, the bottom-right panel shows our reproduction of models from the Sonora Bobcat suite \citep{Marley2021} with [M/H]$ = -0.5$ (corresponding to $Z = 0.00484$) using their boundary conditions\footnote{The Sonora Bobcat boundary tables extend down to $T_{\rm eff} = 200$ K. In this study, we extrapolated the tables below this limit, as did \citet{Marley2021} for the original Sonora models (Mark Marley \& Didier Saumon, private communication)}. The Sonora Bobcat grid solved the coupled atmosphere–interior structure equations over $T_\mathrm{eff} = 200$–$2400$~K and $\log g = 3$–$5.5$, incorporating updated opacities, rainout chemistry, and the SCvH95 EOS with metals. Similar to \cite{Phillips2020} models, the effect of heavy elements in the Sonora models is also not captured through a separate EOS, but is instead mimicked by adopting an effective helium fraction $Y^{\prime} = Y + Z = 0.27834$ (where $Y = 0.2735$). We thus demonstrate that, under the different model assumptions described above, \texttt{APPLE} can successfully reproduce and emulate previous evolutionary calculations.  \textcolor{black}{Table \ref{tab:model_physics_summary} summarizes the input physics and key assumptions adopted in the evolutionary model suites considered in this work. }

\begin{table*}
\centering
\caption{Summary of input physics adopted in reproduced evolutionary model suites in Section \ref{sec:reproduction}.}
\label{tab:model_physics_summary}
\scriptsize
\setlength{\tabcolsep}{4pt}
\begin{tabular}{lcccccc}
\hline\hline
Model suite
& H-He EOS
& Z treatment
& Core
& Atmosphere boundary
& Metallicity / composition
& Other assumptions \\
\hline
\citet{Burrows1997}
& SCvH95
& No Z
& No
& Isolated, non-gray
& $Y=0.25, Z=0$
& Adiabatic; homogeneous; no He rain \\

\citet{Fortney2011}
& SCvH95
& AQUA
& Yes
& Irradiated tables
& $Z=0.059$ (Jup), $0.03$ (Sat)
& Adiabatic; homogeneous; no He rain \\

\citet{Phillips2020}
& CMS19
& $Y' = Y+Z$
& No
& ATMO grids
& $Y' = 0.2919$
& Adiabatic; homogeneous; no He rain \\

\citet{Marley2021}
& SCvH95
& $Y' = Y+Z$
& No
& Sonora-Bobcat grids
& $Z=0.00484$, $Y=0.2735$
& Adiabatic; homogeneous; no He rain \\
\hline
\end{tabular}
\end{table*}

\section{Improvements in Giant Exoplanets Modeling}
\label{sec:improvements}
Having reproduced the heritage evolutionary models using \texttt{APPLE}, we present in this section the individual effects of new aspects that have emerged subsequently in giant planet models and modeling. These include using realistic heavy-element equations of state instead of an augmented helium fraction, improvements in the overall equation of state of hydrogen/helium mixtures, updates to the atmosphere boundary conditions, the effects of helium rain, the effect of the possible presence of an extended fuzzy core of heavy elements, and an example of the potential effects of the initial thermal profiles on the subsequent evolution. \textcolor{black}{We compare our evolutionary tracks primarily to the Sonora-Bobcat emulators as a representative benchmark, with the underlying assumptions and input physics explicitly stated in each case. Throughout this work, we model heavy elements using the AQUA equation of state \citep{Haldemann2020}. Although its original implementation contained an error in the entropy calculation, we find that its impact on the resulting evolutionary models is at the sub-percent level, as demonstrated in the Appendix.}

\begin{figure*}
    \centering
    \includegraphics[width=0.95\linewidth]{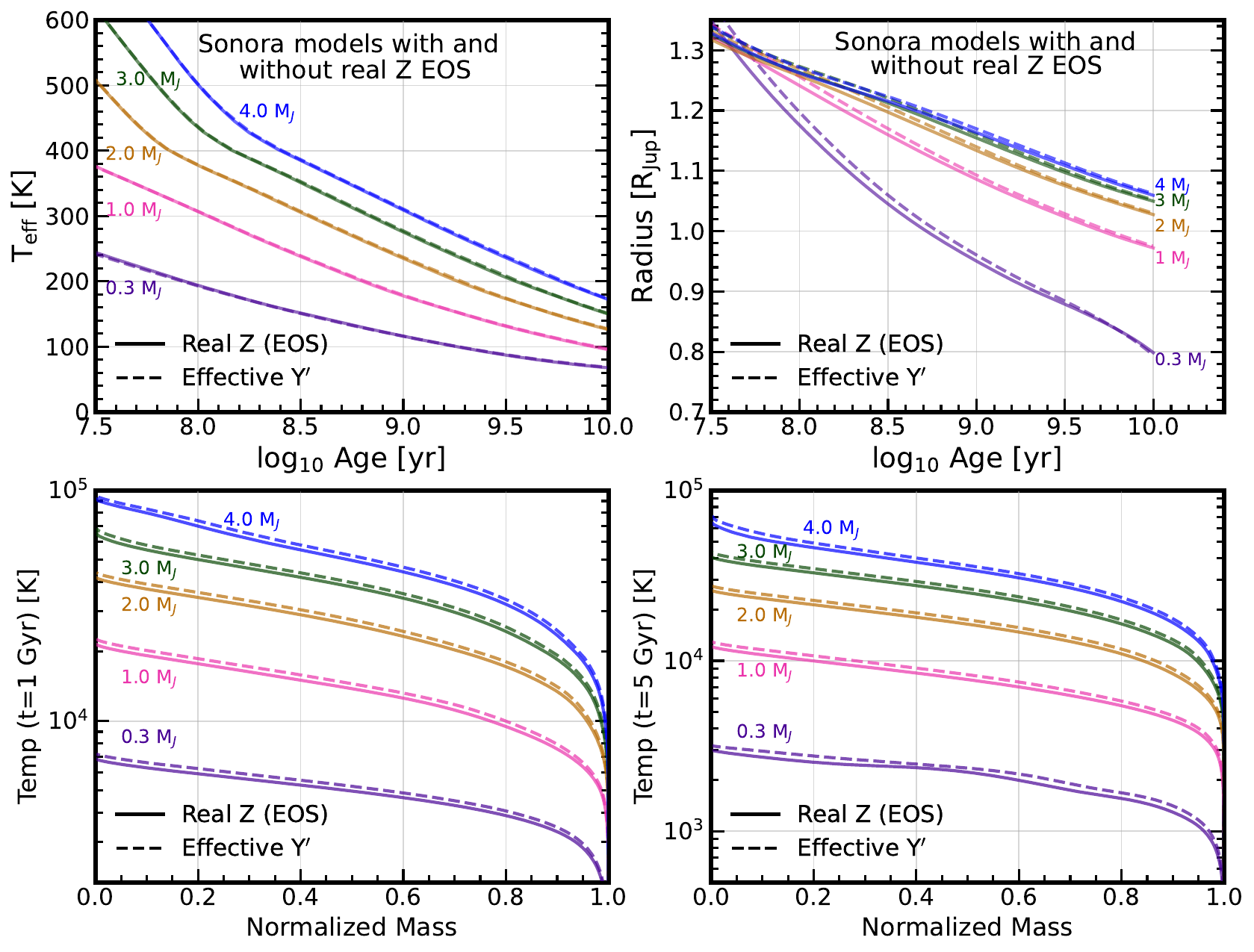}
    \caption{\textcolor{black}{Comparison of adiabatic evolution models of giant planets with different treatments of heavy elements: solid lines correspond to models where heavy elements in the envelope are treated with the AQUA EOS \citep{Haldemann2020}, while dashed lines represent models in which the effect of heavy elements is not modeled with a dedicated $Z$ EOS, but is instead approximated by an effective helium fraction $Y^{\prime} = Y + Z = 0.27834$ (as has traditionally been done). The top panel shows the evolution of the effective temperatures and radii, while the bottom panel shows the temperature profiles at 1 and 5 Gyr, respectively. All calculations use the Sonora Bobcat atmospheric boundary conditions \citep{Marley2021} with [M/H] $= +0.5$, and cover planetary masses ranging from $0.3$ to $4\,M_{\rm Jup}$. For the various masses, the median relative errors range from $\sim 0.6\%$ to $0.75\%$ for $T_{\rm eff}$, $\sim 0.3\%$ to $0.5\%$ for the radii, and $\sim 4.5$ to $7\%$ for the temperature profiles. }}
    \label{fig:yquiv}
\end{figure*}

\begin{figure*}
    \centering
    \includegraphics[width=1.0\linewidth]{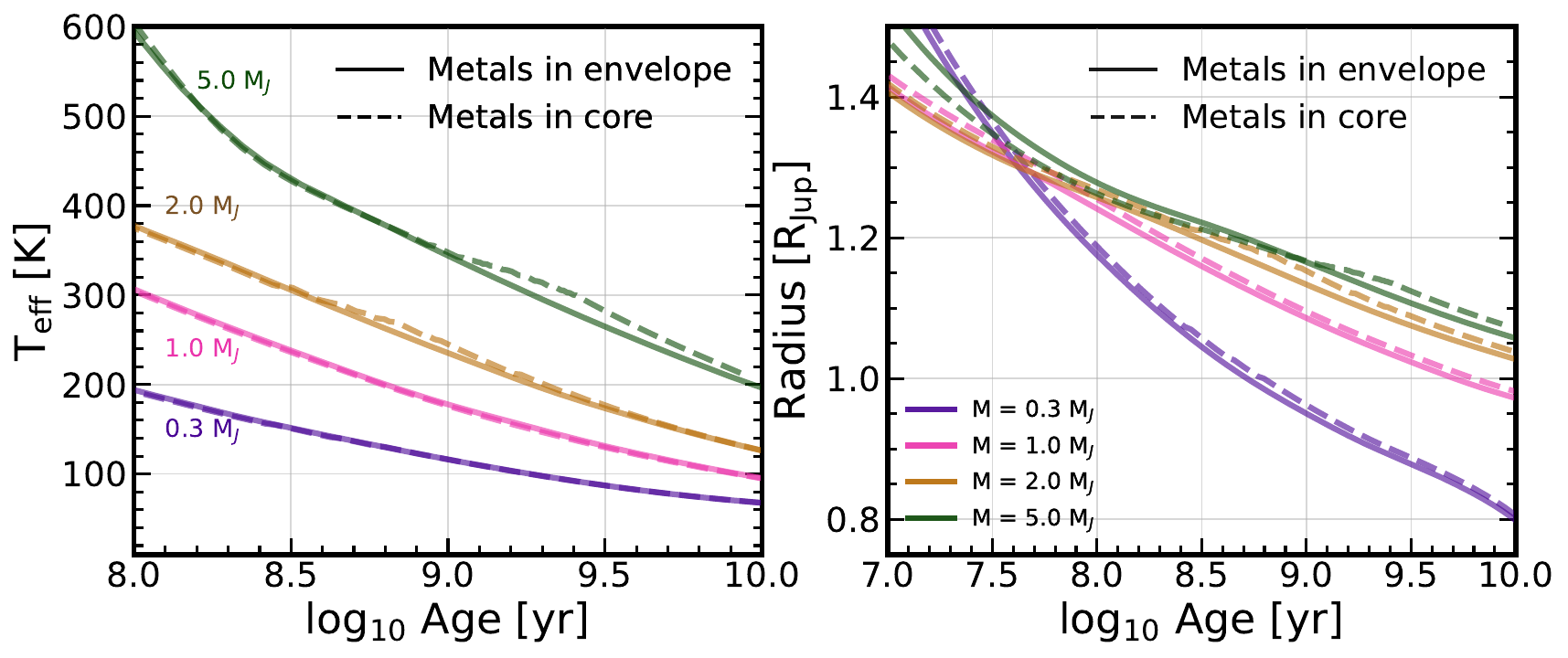}
    \caption{\textcolor{black}{Evolution of giant planets comparing models where (i) metals are uniformly mixed throughout the envelope without a compact core and (ii) all metals are concentrated entirely in the core. The bulk metallicity for all the models is 3.16 $Z_\odot$  ($Z = 0.0484$). All calculations adopt the Sonora Bobcat atmospheric boundary conditions \citep{Marley2021}. Assuming that the same heavy elements reside in the core versus in the envelope results in a difference of $\sim 0.9$ to 1.3\% in the radius for all times. The median relative errors for $T_{\rm eff}$ range from $0.12\%$ for 0.3 M$_{\rm Jup}$ to $5\%$ for 5\% M$_{\rm Jup}$ over 5 Gyr. This suggests that, contrary to the common simplifying assumption of radius being independent of the heavy-element distribution at fixed mass fraction, modest but systematic differences can arise even in adiabatic models.
    }}
    \label{fig:core_env_Z_comparison}
\end{figure*}

\subsection{Treatment of Heavy Elements in Giant Exoplanet Evolution Models}
\label{heavy}
In earlier evolutionary models, heavy elements were often not treated with a dedicated equation of state. Instead, their contribution was approximated by folding them into an effective helium mass fraction, $Y^{\prime} = Y + Z$, as in \citet{Chabrier1997, Chabrier2003, Saumon2008, Phillips2020, Marley2021}. This approach provided a computationally simpler alternative, but neglected the distinct thermodynamic behavior of heavy elements/``metals". \textcolor{black}{\citet{Baraffe2008} were the first to demonstrate the importance of using a realistic EOS for heavy elements when modeling planetary evolution, building on earlier studies that employed ANEOS to represent metals \citep[e.g.,][]{Fortney2006, Fortney2007}. They showed that, for envelope metal mass fractions $Z \lesssim 20\%$, the effects of heavy elements can be approximated by an effective helium fraction $Y^{\prime}$. However, this approximation breaks down at higher metallicities, leading to increasingly inaccurate evolutionary predictions. We reassess the impact of this approximation on the adiabatic evolution of giant planets, comparing such models to those in which envelope heavy elements are treated explicitly using the AQUA EOS \citep{Haldemann2020}.}

The resulting evolutionary tracks are shown in Figure \ref{fig:yquiv} for masses ranging from $0.3$ to $4\, M_{\rm Jup}$. All calculations employ the Sonora Bobcat atmospheric boundary conditions with an envelope metallicity of 3.16 $Z_{\odot}$\citep{Marley2021}, a fixed helium mass fraction $Y=0.2735$ treated with the SCvH95 EOS, and assume coreless, non-rotating planets. Solid lines represent models that employ the AQUA EOS, while the dashed curves specifically correspond to Sonora models that approximate heavy elements with an effective helium fraction of $Y^{\prime} = 0.3219$. In the top panel, we compare the effective temperatures and planetary radii, while in the bottom panel, we show the temperature profiles at 1 and 5 Gyr, respectively, highlighting the differences that arise in the planetary interiors due to the different EOS treatments. 

Quantitatively, the two approaches yield similar evolutionary tracks, with median relative differences of only $\sim0.6$ to $0.75\%$ in effective temperature and $\sim0.3$ to $0.5\%$ in planetary radii. However, the internal temperature profiles show somewhat larger discrepancies, on the order of $4.5$ to $7\%$, underscoring the importance of an explicit EOS treatment when precise thermal structures are required.

\begin{figure*}
    \centering
    \includegraphics[width=1.0\linewidth]{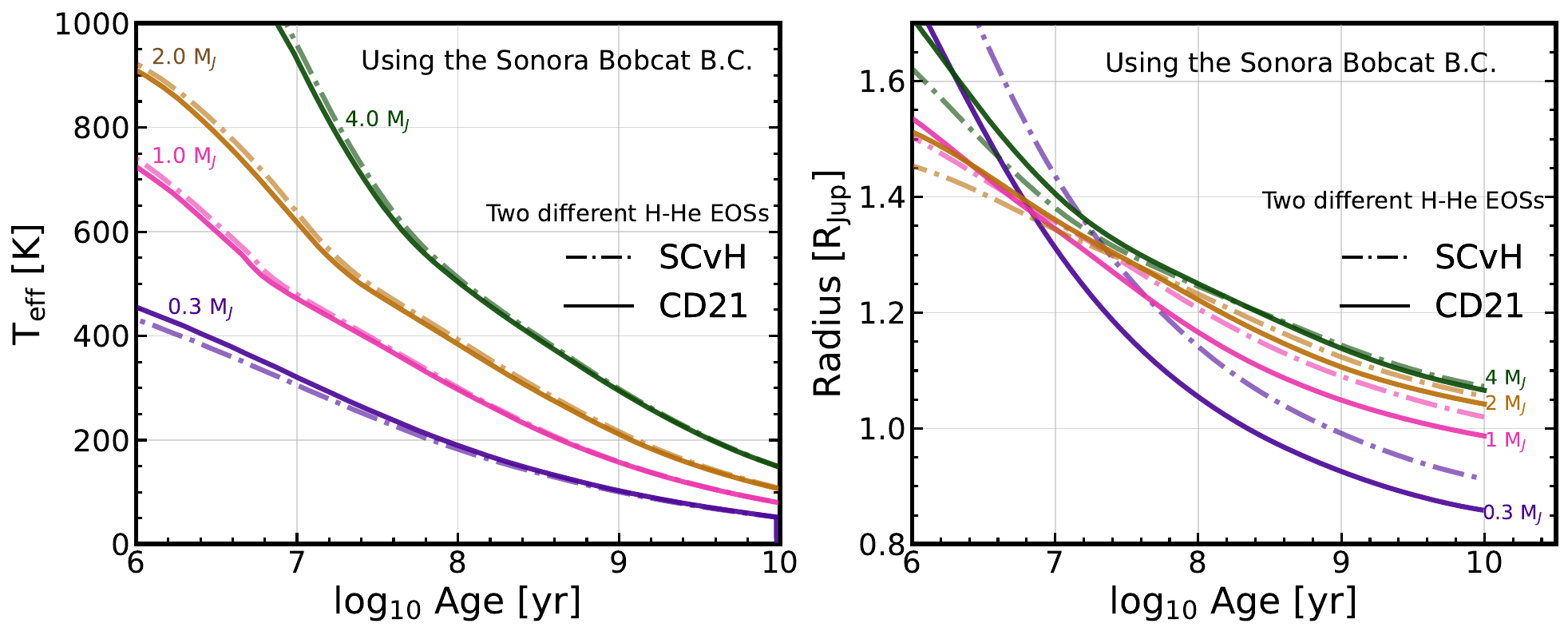}
    \caption{Adiabatic evolution of giant planets, comparing two hydrogen–helium equations of state: the old SCvH95 \cite[dot–dashed lines]{Saumon1995} and the latest CD21 \cite[solid lines]{Chabrier2021} using atmospheric boundary conditions from the Sonora Bobcat grids. The left panel shows the evolution of effective temperature, while the right panel shows the evolution of planetary radius, for masses ranging from $0.3$ to $4\,M_{\rm Jup}$. The median relative errors over 10 Gyr between the two EOSs for $T_{\rm eff}$ range from $1.9\%$ for 0.3 M$_{\rm Jup}$ to $0.6\%$ for 4 M$_{\rm Jup}$. The effect of the modern EOS on radius is more pronounced, increasing from $\sim 0.6\%$ at higher masses to $\sim 6.7\%$ at lower masses.}
    \label{fig:eos_comp}
\end{figure*}
\subsection{Impact of Metal Distribution on Evolutionary Properties}

We next examine how the spatial distribution of heavy elements influences the evolution of giant planets. Two limiting configurations are considered: in the first, heavy elements are assumed to be homogeneously mixed throughout the entire envelope, with no distinct compact core. In the second, the same heavy-element mass is concentrated entirely into a central core, leaving a metal-free envelope. \textcolor{black}{This issue was previously examined by \citet{Baraffe2008}, who showed that for total heavy-element mass fractions $Z \lesssim 10$–15\%, the evolutionary effects of metals can be well approximated by concentrating all heavy elements in the core. At higher metallicities ($Z \gtrsim 20\%$), however, cooling histories and radius predictions diverge by up to $\sim$10\% at a given age. We revisit this regime by considering models with a total heavy-element mass fraction} of $Z=0.0484$ (3.16 $Z_\odot$), which corresponds to heavy-element masses of approximately 4.61, 15.38, 30.76, and 61.52 $M_\oplus$ for planets of 0.3, 1, 2, and 4 $M_{\rm Jup}$, respectively. 

All evolutionary calculations employ the Sonora Bobcat atmospheric boundary conditions with $Z=0.0484$ \citep{Marley2021}. The hydrogen–helium component of the envelope is treated with the SCvH95 EOS for a fixed helium fraction $Y=0.2735$, while the heavy elements are represented using the AQUA EOS \citep{Haldemann2020}, both in the envelope and in the core. This consistent treatment allows for a controlled comparison between the two distributional exemplars.

The results are shown in Figure \ref{fig:core_env_Z_comparison} for both the time evolution of the effective temperature and planetary radius, and the relative differences between the two cases. While the global properties remain broadly similar, the models reveal systematic differences: planets with metals distributed throughout the envelope are systematically smaller than planets with metals concentrated in the core. The radius differences are on the order of  $\sim1.0–1.3\%$ across all masses and evolutionary times, while effective temperatures vary by only $\sim$0.7–1\%. These results suggest that even in adiabatic models, the distribution of heavy elements can introduce measurable shifts, complementing the more commonly examined dependence on total heavy-element content.

\subsection{Comparison of Hydrogen–Helium Equations of State}

All heritage exoplanet models, such as \citet{Burrows1997, Chabrier1997, Baraffe1998, Baraffe2003, Fortney2004, Saumon2008, Marley2021}, used the SCvH95 \citep{Saumon1995} EOS tables for modeling hydrogen and helium. To quantify the influence of hydrogen–helium thermodynamics on planetary evolution, we compare the older SCvH95 model to the more recent CD21 tables \citep{Chabrier2021}. \textcolor{black}{These EOSes differ primarily in their treatment of non-ideal effects and phase behavior: in particular, \citet{Chabrier2019} revised the isothermal compressibility of hydrogen at the high temperatures and pressures relevant to gas-giant interiors. This comparison therefore provides a direct test of how updates in microscopic physics propagate into macroscopic evolutionary observables of giant planets.}

We evaluate the models under the Sonora Bobcat  \citep{Marley2021} atmospheric boundary conditions as implemented within the \texttt{APPLE} code. The left panel of \ref{fig:eos_comp} represents the evolution of effective temperatures, while the right panel shows the evolution of planetary radii for masses ranging from $0.3$ to $4\,M_{\rm Jup}$. The models are isolated and coreless, with an envelope metallicity of $Z=0.00484$\footnote{\textcolor{black}{It corresponds to [Fe/H]= -0.5, chosen in order to minimize the influence of the heavy-element EOS on the volume addition law when comparing different H–He EOSs.}} and helium mass fraction $Y=0.2735$. In these models, heavy elements are not treated with a dedicated EOS, but are instead folded into an effective helium fraction, $Y^{\prime} = Y+Z = 0.27834$, following the prescription of \citet{Marley2021}.

The quantitative differences between SCvH95 and CD21, shown in Figure \ref{fig:eos_comp}, depend primarily on planetary mass. For effective temperatures, the relative differences between the two EOSs remain modest, ranging from $\sim1.9\%$ for the $0.3\, M_{\rm Jup}$ case to $\sim0.6\%$ for the $4\,M_{\rm Jup}$ case. The impact on planetary radii, however, is more pronounced. Over 10 Gyr, the radius differences increase from $\sim0.6\%$ at higher masses to as much as $\sim6.7\%$ at lower masses. These results indicate that while updates to the hydrogen–helium EOS have only a limited effect on the thermal evolution of more massive giant planets, they can lead to significant structural differences for lower-mass giants. Similar conclusions were reached by \citet{Chabrier2021} and \citet{Howard2025b}, the latter comparing results of the \citet{Chabrier2019} H-He EOS with the \citet{Howard2023a} corrections to the SCvH95 EOS on giant exoplanet interiors. Further, it has also been shown that the difference in H-He EOS for Jupiter and Saturn has consequences for superadiabaticity \citep{Howard2024}, and the total metal content required to fit the Juno data \citep{Miguel2016}.

\begin{figure*}
    \centering
    \includegraphics[width=1.0\linewidth]{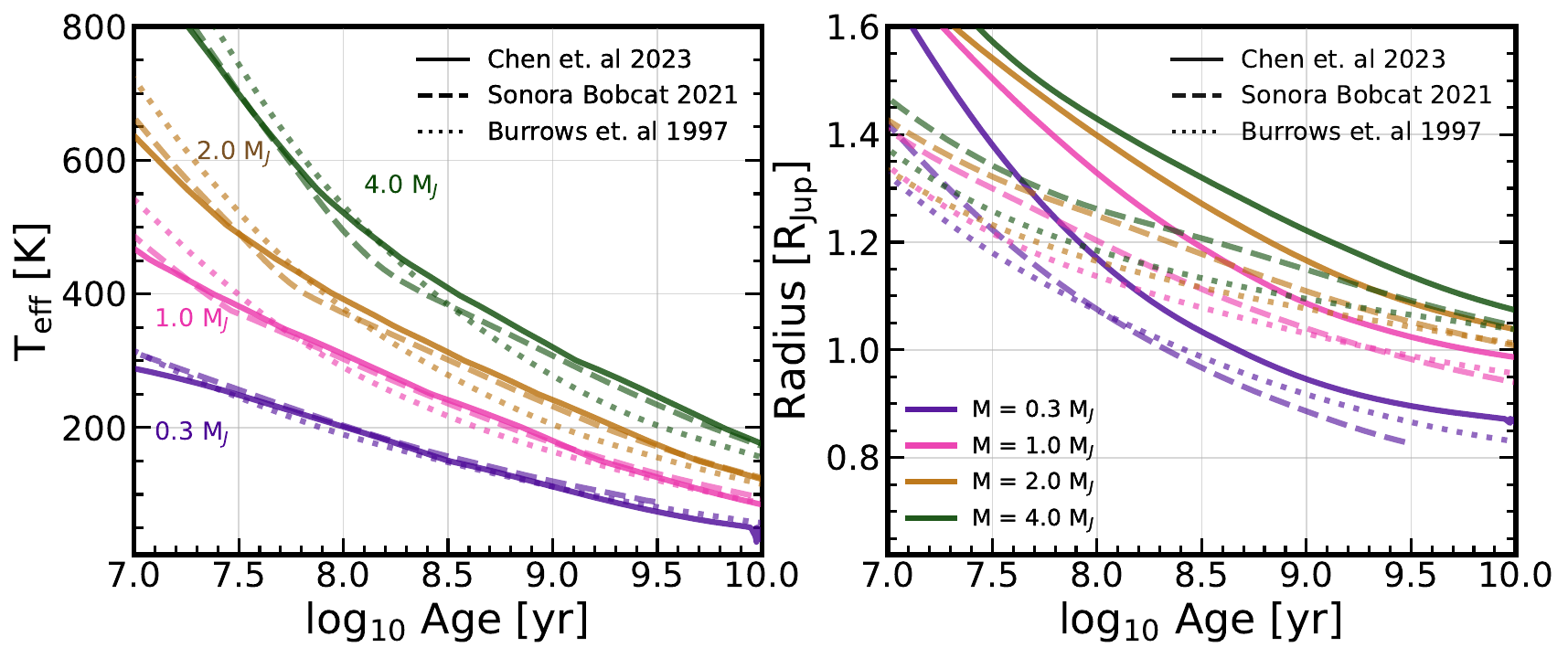}
    \caption{\textcolor{black}{Adiabatic evolution of giant planets comparing three atmospheric boundary conditions: \citet[][solid lines]{Chen2023}, \citet[][Sonora-Bobcat, dashed lines]{Marley2021}, and \citet[][dotted lines]{Burrows1997}. Models are shown for planetary masses ranging from $0.3$ to $4\,M_{\rm Jup}$. All calculations assume coreless planets with homogeneous and adiabatic hydrogen–helium envelopes mixed with heavy-element ices, treated using the AQUA EOS. At late times, the effective temperature differences among the various atmospheric boundary conditions range from about $7\%$ for lower-mass planets to $\sim 3\%$ for higher-mass cases. The corresponding radii differ by approximately $3$ to $6\%$ at late times, while at early times the discrepancies can be significantly larger.}}
    \label{fig:boundary_conditions_comparison}
\end{figure*}

\subsection{Sensitivity to Atmospheric Boundary Conditions}

We also investigate how the choice of atmospheric boundary conditions influences the adiabatic evolution of giant planets. Two widely used boundary models: the grids of the Sonora Bobcat models of \citet{Marley2021}, and the earlier tables of \citet{Burrows1997}, are compared with the more recent \citet{Chen2023} boundary condition. The latter was originally developed to model Jupiter and Saturn \citep{Tejada2024, Sur2025a, Sur2025b} and included irradiation, but we have expanded our isolated boundary tables to cover the giant exoplanet parameter space. At the lowest effective temperatures, the \citet{Chen2023} boundary conditions also include the effects of ammonia cloud formation. Hence, these various outer thermal boundary models represent different generations of atmosphere calculations, incorporating advances in opacities, radiative transfer, and molecular chemistry, and thus provide a benchmark for the robustness of planetary cooling models to outer boundary assumptions. 

Our calculations span planetary masses from $0.3$ to $4\,M_{\rm Jup}$. All models shown in this section assume coreless planets with fully adiabatic, homogeneous hydrogen–helium envelopes with envelope metallicity of 3.16 $Z_{\odot}$. The helium mass fraction is fixed at $Y=0.2735$, treated with the SCvH95 EOS, while heavy elements are modeled as ices using the AQUA EOS. This setup isolates the effect of the atmosphere tables by keeping the interior physics fixed.

The results shown in Figure \ref{fig:boundary_conditions_comparison} reveal that boundary conditions play a non-negligible role in shaping the thermal evolution. At late times, effective temperatures differ by $\sim 7\%$ for the $0.3\,M_{\rm Jup}$ case, while the discrepancy decreases to $\sim 3\%$ for the $4\,M_{\rm Jup}$ planet. The corresponding planetary radii at a given age differ by about 3–6\% across the mass range. Early in the evolution, however, the spread among models is much larger, reflecting both the stronger influence of boundary conditions at high entropies and the critical uncertainty in the initial thermal profiles inherited from formation.

Overall, these comparisons demonstrate that atmospheric boundary conditions introduce systematic uncertainties in both $T_{\rm eff}$ and planetary radii, especially at young ages, and highlight the importance of using updated atmosphere models in conjunction with consistent initial conditions.

\begin{figure*}
    \centering
    \includegraphics[width=1.0\linewidth]{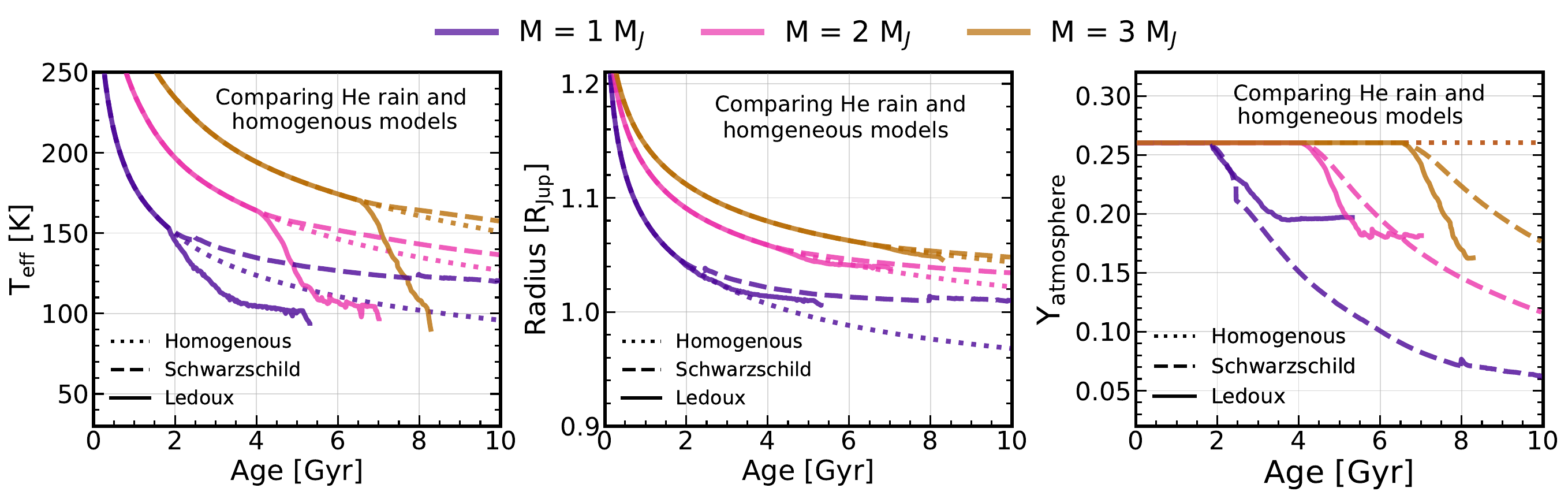}
    \caption{\textcolor{black}{Evolution of giant planets comparing helium-rain models with homogeneous adiabatic Sonora Bobcat models without it for $1$, $2$, and $3\,M_{\rm Jup}$ planets. Solid curves indicate convective stability using the Ledoux criterion, while dashed curves correspond to the Schwarzschild criterion. All models employ the Sonora Bobcat atmospheric boundary conditions with 3.16 $Z_{\odot}$ metallicity. The hydrogen–helium miscibility curve of \citet[L0911]{Lorenzen2009, Lorenzen2011}, shifted by $+300$ K, is adopted in this study. At 5 Gyr, helium-rain models differ from homogeneous cases by $-20$ to $-60$ K in $T_{\rm eff}$ under Ledoux convection and by $+25$ to $+5$ K under Schwarzschild convection, with radius offsets of only $\sim1$ to $3\%$. Atmospheric helium is moderately depleted in Ledoux models ($Y_{\rm atm} \sim0.16$ to 0.20) but strongly depleted in Schwarzschild models ($Y_{\rm atm} \sim0.06$ by 10 Gyr).}}  
    \label{fig:helium_rain_comparison}
\end{figure*}
\subsection{Effects of Helium Rain on Giant Exoplanet Evolution}

Helium rain has long been recognized as a key process in the evolution of solar system giant planets, invoked originally to explain the depleted atmospheric helium abundance in Jupiter and to account for the anomalous luminosity of Saturn \citep{Stevenson1977b, Fortney2003, Mankovich2016, Pustow2016, Mankovich2020, Howard2024, Tejada2025, Sur2025a}. To date, the only study that has explicitly considered helium rain in the context of exoplanets is \citet{Fortney2004}. Given its demonstrated importance for the solar system giants, incorporating helium rain should be a priority for the next generation of exoplanet evolutionary models, particularly at the lowest masses.

To explore the role of helium phase separation, we compare evolutionary models with and without helium rain for planets of 1, 2, and $3,M_{\rm Jup}$. The baseline reference is given by homogeneous, fully adiabatic Sonora Bobcat models, against which we contrast helium-rain models calculated with the diffusion scheme B implemented in \texttt{APPLE} \citep{Sur2024a}. Convective stability is assessed using both the Ledoux and Schwarzschild criteria, as shown by the solid and dashed curves, respectively, in Figure \ref{fig:helium_rain_comparison}.

All models adopt the Sonora Bobcat atmospheric boundary conditions at 3.16 $Z_{\odot}$ metallicity. Hydrogen–helium mixtures are described with the SCvH95 EOS, while helium immiscibility is governed by the phase diagram of \citet[L0911]{Lorenzen2009, Lorenzen2011}, shifted upward by $+300$ K. This calibration was previously demonstrated to reproduce both Jupiter’s and Saturn’s effective temperatures and atmospheric helium abundances \citep{Tejada2024, Sur2025b}, and we therefore apply the same shift here. Neither rotation nor the presence of a dense core is considered in this set of calculations.

The results highlight moderate, but systematic, differences between homogeneous and helium-rain cases for the Sonora Bobcat models. At 5 Gyr, the effective temperatures of helium-rain models differ from homogeneous models by approximately $-20$ K to $-60$ K under Ledoux convection, and by $+25$ to $+5$ K under Schwarzschild convection, from lower- to higher-mass planets, respectively. This contrast has not been explored previously by \citet{Fortney2004}, who considered only Schwarzschild convection and, therefore, found only a rise in effective temperature during evolution. The median relative error in planetary radius at this age is $\sim 1\%$ to $3\%$ for both convection criteria. The atmospheric helium fraction in the Ledoux models spans $Y_{\rm atm} \sim 0.16$ to $0.20$, whereas the Schwarzschild models exhibit much stronger depletion, reaching as low as $Y_{\rm atm} \sim 0.06$ at 10 Gyr. Note that the $3 \, M_{\rm Jup}$ planet undergoes helium rain only at much later times ($>6$ Gyr) and remains nearly adiabatic, indicating that helium rain plays a diminishing role in more massive exoplanets. These calculations do not attempt to fit Jupiter and exclude both fuzzy cores and inhomogeneous metal distributions, and therefore yield different atmospheric helium abundances than our previous predictions \citep{Sur2025a, Sur2025b}. We further note that the Ledoux models become noisy at low temperatures ($\sim 100$ K), where the $T_{\rm eff}$s are significantly extrapolated beyond the Sonora Bobcat boundary condition tables\footnote{private communication: Mark Marley \& Didier Saumon}. \textcolor{black}{These particular models are truncated due to numerical instabilities encountered for that one model, and do not extend to the full evolution}. Overall, we conclude that helium rain plays a significant role in the early evolutionary stages of lower-mass gas giants, altering their cooling rates, but its influence diminishes progressively for planets more massive than $\sim3\,M_{\rm Jup}$ at Solar System ages.

\begin{figure*}
    \centering
    \includegraphics[width=0.95\linewidth]{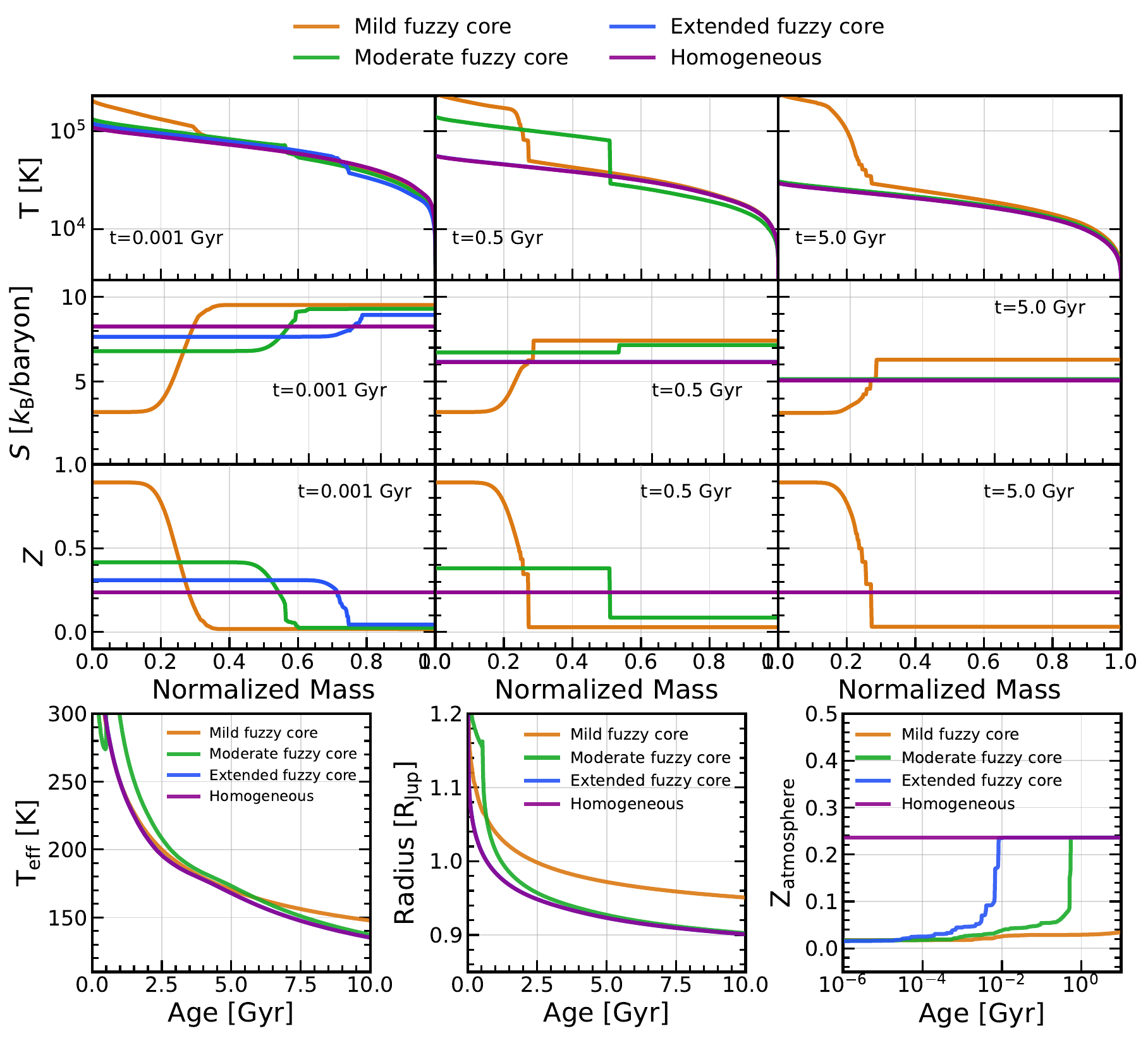}
    \caption{\textcolor{black}{Comparison of the evolution of a $2\,M_{\rm Jup}$ exoplanet with three different initial fuzzy-core extents—mild, moderate, and extended—modeled with the AQUA EOS, against a homogeneous adiabatic Sonora-Bobcat model. The helium fraction is fixed at $Y=0.2735$ (CD21 EOS), and all models adopt Sonora-Bobcat atmospheric boundary conditions at $3.16\,Z_{\odot}$. The bulk metallicity is $Z=0.236$, corresponding to a heavy-element mass of $150\,M_{\oplus}$, chosen to allow substantial mixing and to test evolutionary differences between fuzzy-core configurations. The top panels show the interior temperature, entropy, and heavy-element fraction profiles at 0.001, 0.5, and 5 Gyr, while the bottom panels track the evolution of effective temperature, radius, and atmospheric metallicity. All fuzzy-core models begin from nearly identical temperature profiles with hot-start conditions. Relative to the homogeneous model, the fuzzy-core cases show median deviations of $8\%$ to $12\%$ in $T_{\rm eff}$ and $8\%$ to $13\%$ in radius. The atmospheric metallicity rises from $Z=0.02$ (mild) to $Z=0.25$ (extended).}}
    \label{fig:fuzzy_core_extent}
\end{figure*}

\subsection{Influence of Fuzzy Cores on Giant Exoplanet Evolution}

Results from Juno \citep{Bolton2017a} and Cassini \citep{Iess2019} have fundamentally reshaped interior models of the solar system’s giant planets, favoring the presence of ``fuzzy" cores \citep{Wahl2017, Militzer2022, Militzer2024, Mankovich2021} over traditional compact ones. Such fuzzy cores have already been incorporated into evolutionary models of Jupiter and Saturn \citep{Tejada2025, Sur2025a, Sur2025b}, but they have not yet been systematically explored in the context of exoplanets. An important open question concerns both the initial extent of the fuzzy core and its long-term survival over Gyr timescales, since convective mixing is expected to gradually erode compositional gradients \citep{Knierim2024, Tejada2025, Fuentes2021, Fuentes2025}, potentially leading to full homogenization. This not only has observational consequences for their atmospheric metallicity \citep{Knierim2025}, but also for the origin of close-in giant planets \citep{Knierim2022}.

We investigate the evolutionary consequences of different initial fuzzy-core extents by comparing models of a $2\,M_{\rm Jup}$ exoplanet with three distinct degrees of fuzzy core extent: mild, moderate, and extended. All fuzzy-core models are constructed using the AQUA EOS for heavy elements, with hydrogen–helium mixtures treated using the CD21 EOS. The helium mass fraction is fixed at $Y=0.2735$, and the models employ Sonora Bobcat atmospheric boundary conditions at 3.16 $Z_{\odot}$. The comparison also includes a homogeneous adiabatic reference model, in which the envelope metallicity is $Z=0.236$. All models have a total heavy element mass of 150 $M_{\odot}$. This is chosen as an example to illustrate the models' ability to undergo substantial mixing, allowing us to quantify the differences between various observables in this context. Rotation and helium rain are neglected here to isolate the role of fuzzy cores.

The evolutionary tracks are shown in terms of both internal and global properties. The top panels of Figure \ref{fig:fuzzy_core_extent} illustrate the evolution of temperature, entropy, and heavy-element fraction profiles at three characteristic ages: 0.001 Gyr, 0.5 Gyr, and 5 Gyr. The bottom panels present the corresponding time evolution of effective temperature, planetary radius, and atmospheric metallicity. The results demonstrate that the extent of the initial fuzzy core strongly influences the efficiency and timescale of compositional mixing. The extended fuzzy core rapidly homogenizes with the envelope ($\leq 0.01$ Gyr), followed by the moderately fuzzy core ($\sim 1 $ Gyr), while the mildly fuzzy core remains largely intact and persists beyond 10 Gyr. The extent of mixing is, however, dependent on the chosen initial entropy profiles, as hotter entropies will lead to rapid homogenization of the fuzzy cores \citep{Knierim2024, Tejada2025}. 

Quantitatively, the fuzzy-core models differ appreciably from the homogeneous case. The median relative deviations range from $8\%$–$12\%$ for effective temperature and $8\%$–$13\%$ for planetary radius, indicating that observable properties retain a measurable memory of the planet’s initial compositional structure. The atmospheric metallicity also evolves in a manner correlated with the degree of initial ``fuzziness", increasing from $Z \simeq 0.02$ in the mild fuzzy case to $Z \simeq 0.25$ in the extended and moderate fuzzy cases. These results highlight the importance of fuzzy-core physics \citep{Fuentes2025} in shaping both the long-term cooling and the observable atmospheric properties of giant exoplanets.

\begin{figure*}
    \centering
    \includegraphics[width=1.0\linewidth]{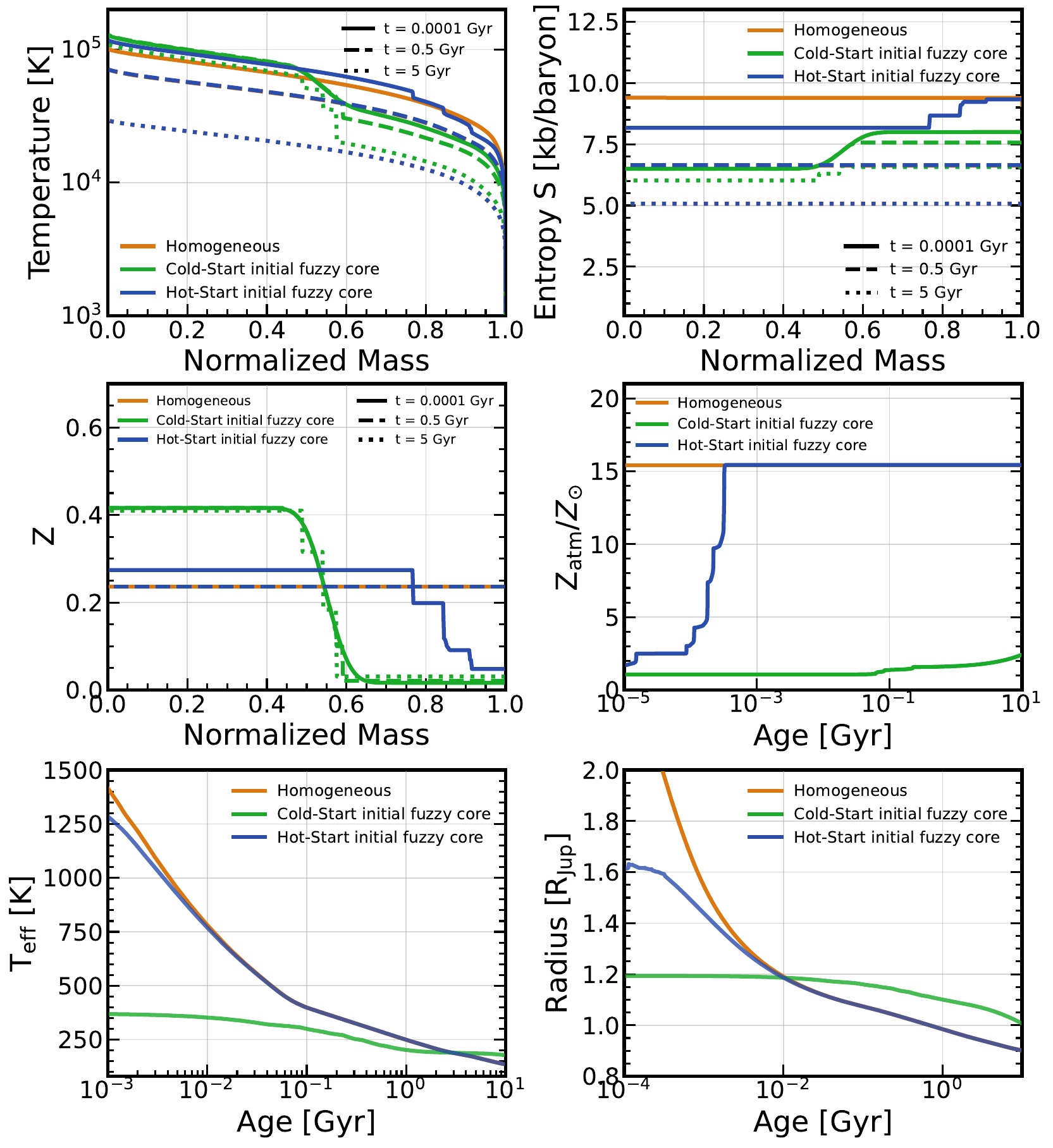}
    \caption{\textcolor{black}{Comparison of evolutionary models for a $2\,M_{\rm Jup}$ planet with different initial conditions: a hot-start initial fuzzy-core model (blue lines), a cold-start initial fuzzy-core model (green lines), and a homogeneous Sonora Bobcat model (orange lines). The initial central temperatures for the hot fuzzy core model and the cold fuzzy core model are $1.5\times10^5$ K and $1.2\times10^5$ K; however, their outer entropies are 9 and 8 $k_B$/baryon, respectively. The homogeneous model starts at 10 $k_B$/baryon. The hot-start fuzzy-core model rapidly homogenizes within $\sim 0.1$ Myr and converges to the homogeneous track, while the cold-start model mixes more slowly, preserving its compositional gradient for over 10 Gyr. At Gyr timescales, the hot-start model shows differences of $<0.5$ K in $T_{\rm eff}$ and $\sim 0.01\%$ in radius compared to the homogeneous case, whereas the cold-start model remains hotter by $\sim 50$ K, with radii larger by $\sim 13\%$ and a correspondingly hotter interior. The atmospheric metallicity also differs: rising to $Z \sim 15\, Z_{\odot}$ in the hot-start case and to $Z \sim 2.5 \,Z_{\odot}$ in the cold-start case.}}
    \label{fig:fuzzy_hot_cold}
\end{figure*}
\subsection{Influence of Initial Conditions on the Fuzzy-Cores of Giant Exoplanet Evolution}
\label{initial_fuzzy)}

Here, we examine how different initial thermal states affect the long-term evolution of a $2\,M_{\rm Jup}$ planet by comparing three models: a hot-start fuzzy-core case, a cold-start fuzzy-core case, and a homogeneous Sonora Bobcat model. The fuzzy-core models begin with extended regions of heavy elements, while the homogeneous model is coreless and assumes a uniform envelope metallicity of $Z=0.236$. The total heavy element mass in all our models is 150 $M_{\oplus}$. All calculations employ the Sonora Bobcat atmospheric boundary conditions at $3.16 \,Z_{\odot}$, the SCvH95 EOS for hydrogen–helium, and the AQUA EOS for heavy elements. For clarity, rotation and helium rain are excluded from this comparison. The initial entropies of the cold- and hot-start fuzzy-core models are $8 \, k_B$/baryon and $9 \, k_B$/baryon, respectively, corresponding to a central temperature difference of $\sim 30{,}000$ K. The homogeneous model is also initialized in a hot-start state.

The results, shown in Figure \ref{fig:fuzzy_hot_cold}, highlight clear differences in both the thermal evolution and structural properties of the models. Over time, the hot-start fuzzy-core model converges to the homogeneous case, with effective temperatures differing by less than $0.5$ K and radii by only $\sim 0.01\%$ (bottom panels). In contrast, the cold-start fuzzy-core model evolves more distinctly, cooling to an effective temperature $\sim 90$ K higher and yielding radii larger by $\sim 13\%$ relative to the homogeneous model. The mixing histories also diverge: the hot-start fuzzy-core model homogenizes within 0.001 Gyr, erasing its initial gradient, whereas the cold-start model mixes more slowly and retains its compositional structure for up to 10 Gyr (middle left panel).

These evolutionary pathways have direct implications for atmospheric properties (see middle right panel). In the hot-start case, the envelope becomes substantially enriched, with atmospheric metallicities rising to $Z \approx 0.23$ by 10 Gyr. In contrast, the cold-start model maintains only modest enrichment, with atmospheric metallicity reaching $Z \approx 0.04$ over the same timescale. Together, these results underscore the sensitivity of planetary evolution to initial thermal states, showing that hot- versus cold-start conditions can imprint long-lasting signatures on both bulk properties and observable atmospheric compositions. We note that 
our best-fit models for Jupiter and Saturn, in which initially fuzzy cores survive to the present day, have low initial internal entropies \citep{Sur2025a,Tejada2025}.

\textcolor{black}{We neglect semiconvective heat transport in the present models, as there is currently no self-consistent treatment of semiconvection in one-dimensional planetary evolution codes. In \citet{Sur2025b}, we explored the evolution of Jupiter and Saturn with fuzzy cores by varying the $R_{\rho}$ parameter, thereby bracketing the limiting transport regimes. Specifically, $R_{\rho}=1$ corresponds to the Ledoux-stable limit, while $R_{\rho}=0$ recovers the Schwarzschild limit with fully efficient convective transport, in which semiconvective layers become fully convective on short timescales \citep{Moll2017Double-diffusiveJupiter,Garaud2018, Tulekeyev2024a}. In reality, semiconvection is expected to enhance heat and compositional transport relative to the Ledoux case, but remain less efficient than fully developed convection \citep{Leconte2012}, such that the resulting evolutionary trajectories lie between these two extremes. We examine the implications of these two limiting regimes in the next section. However, we emphasize that this approach is crude and that a more physically motivated treatment of semiconvection is required to robustly understand the evolution of next-generation giant exoplanets.}

\begin{figure*}
    \centering
    \includegraphics[width=1.0\linewidth]{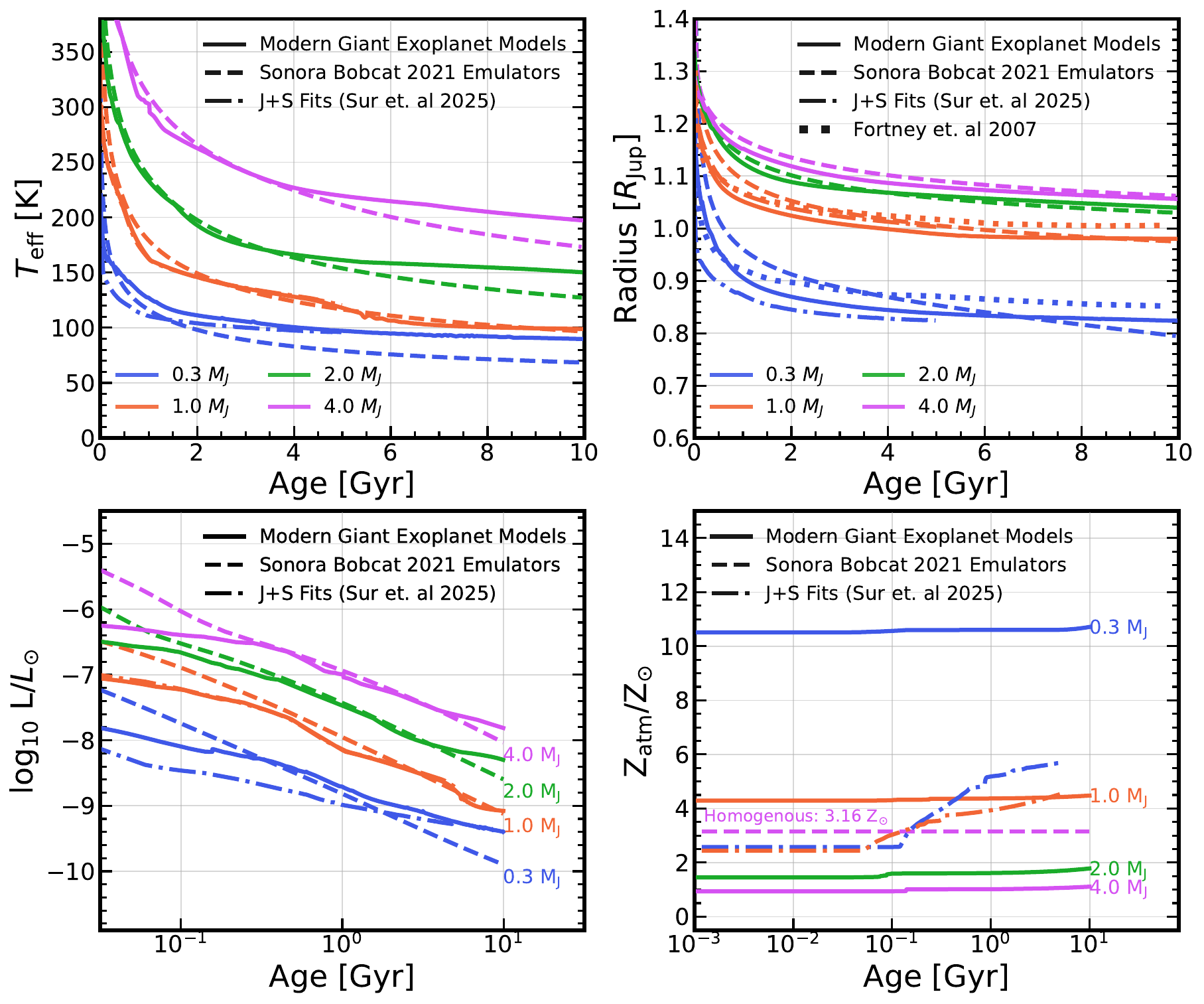}
    \caption{\textcolor{black}{Comparison of sample next-generation evolutionary models for giant exoplanets with updated treatments of heavy-element distributions, EOS, and boundary conditions against the Sonora Bobcat heritage models. Solid lines represent models from this work, dashed lines are Sonora Bobcat emulators \citep{Marley2021}, and dash–dot lines correspond to Jupiter and Saturn fits from \citet{Sur2025a}. We also show in the top-right panel, for comparison, using the dotted line, the radii for 0.3 and 1 M$_{\rm Jup}$ giant planets from \citep{Fortney2007}, which uses a compact rocky core of 25 M$_{\oplus}$.  Models are shown for $0.3$, $1$, $2$, and 4 M$_{\rm Jup}$, corresponding to heavy-element masses of $21$, $42$, $71$, and $128\,M_{\oplus}$. At 5 Gyr, effective temperatures differ from the homogeneous Sonora sequences by $18$, $1.8$, $7.4$, and $8.3$ K for $0.3$, $1$, $2$, and 4 M$_{\rm Jup}$, respectively, while radii differ by only $\sim 2\%$. The \cite{Fortney2007} models, however, predict a much higher radius by up to 5\%. Luminosities agree for higher-mass planets, but the 0.3 M$_{\rm Jup}$ model remains an order of magnitude brighter. Outer metallicities differ significantly, reflecting the influence of interior mixing and helium rain.}}
    \label{fig:final_teff_rad_masses}
\end{figure*}

\section{Sample Next-Generation Evolutions of Giant Exoplanets}
\label{sec:new_models}

Early models of planet formation often assumed that all heavy elements accreted directly to the core. However, more sophisticated formation models and recent measurements of the solar-system giants now suggest that a substantial fraction—if not the majority—of accreted solids are deposited into the envelope through ablation and fragmentation once the core reaches a mass of roughly $0.5$–$3\, M_{\oplus}$ \citep{Inaba2003, Iaroslavitz2007, Hori2011, Browers2018, Valletta2019, Valletta2020}. Determining both the total amount of heavy elements and their distribution within giant exoplanets remains a major challenge, as these properties are highly sensitive to the specific formation pathways \citep{Mordasini2009, Danti2023, Mordasini2015}. Detailed formation models for giant planets are currently available only for Jupiter \citep{Lozovsky2017, Stevenson2022} and Saturn \citep{Bodenheimer2025}, and even in these cases, the origin of their fuzzy cores is still unresolved \citep{Helled2017, Helled2022b, Meier2025, Fuentes2025}. In this section, we assume that all planets begin with fuzzy cores; however, whether these structures survive throughout evolution depends strongly on their initial interior entropy profiles \citep{Knierim2024, Tejada2025, Sur2025a} as well as their extent, as discussed in \S\ref{initial_fuzzy)}. To investigate this, we show the potential effects of both cold- and hot-start initial conditions \citep{Spiegel2012}, and further we compare evolutionary outcomes under the Ledoux and Schwarzschild criteria for convection as done in \cite{Sur2025b}.

\begin{figure*}
    \centering
    \includegraphics[width=1.0\linewidth]{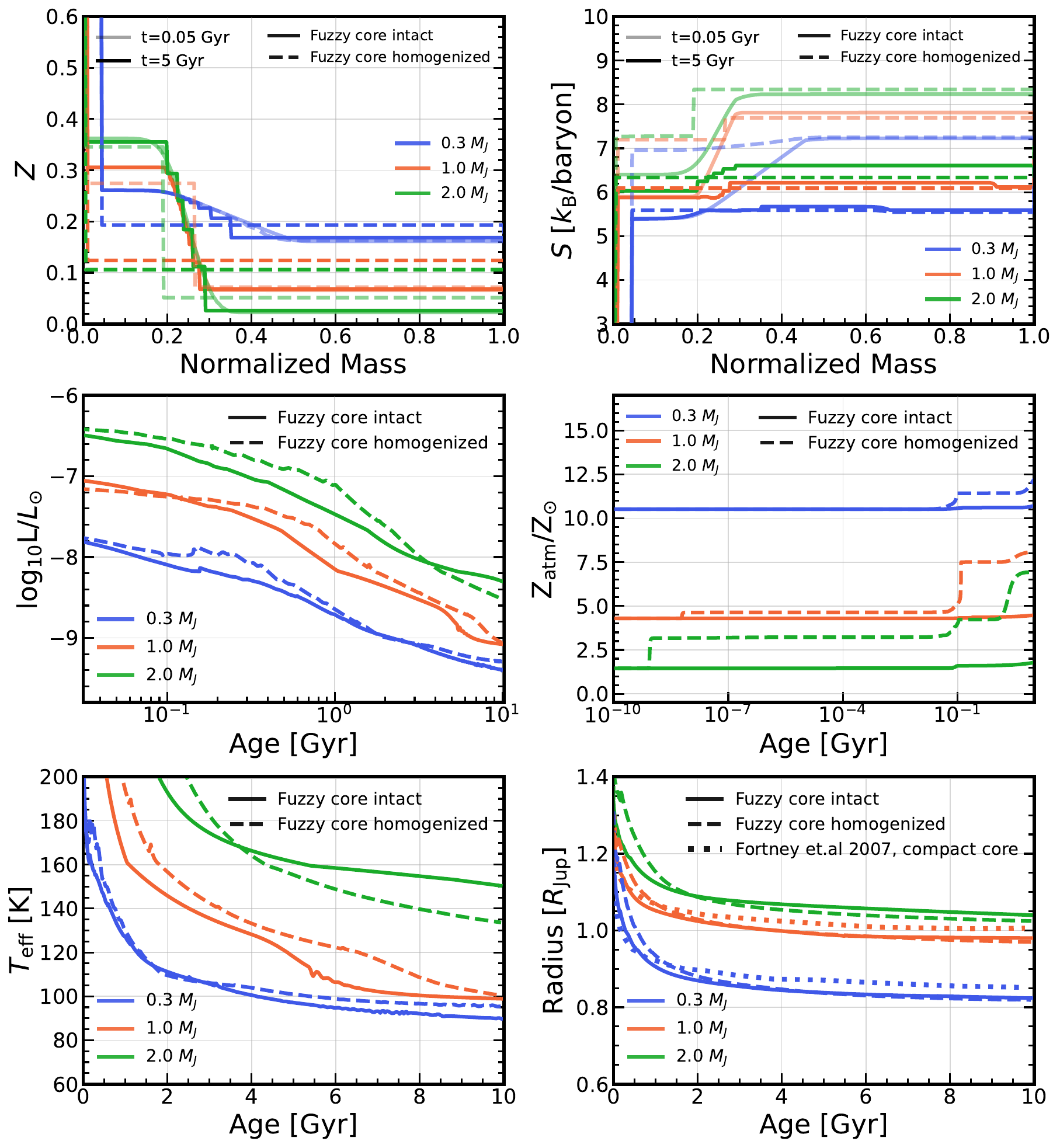}
    \caption{Comparison of next-generation evolutionary models in which fuzzy cores either survive (solid lines) or homogenize (dashed lines) during 10 Gyr of evolution for $0.3\,M_{\rm Jup}$, $1\,M_{\rm Jup}$, and $2\,M_{\rm Jup}$ exoplanets, incorporating the same combined updated physics as in Figure 9. All models begin with an initial fuzzy core extending to 30\% of the total planetary mass, but differ in their initial entropy profiles: cold-start models (interior entropies $\sim$ 6 to $6.5$ k$_B$/baryon), where the fuzzy core survives throughout the evolution, versus hot-start models (interior entropies $\sim 7.5$ to 8.5 k$_B$/baryon), where the fuzzy core rapidly homogenizes. This is demonstrated in the top panels, which show Z and entropy distributions at 0.05 Gyr (translucent lines) and 5 Gyr (dark lines). The middle left panel shows the log of luminosities. In the middle right panel, atmospheric metallicities are shown to have increased by 3 to 5 $Z_{\odot}$ in models that homogenized as a result of substantial mixing. In the cases where the fuzzy core survives, the effective temperatures differ by $-4.0$ K, $-9.1$ K, and $+4.8$ K, while the radii differ by $0.11\%$, $-0.05\%$, and $+1.54 \% $ for $0.3$, $1$, and $2\,M_{\rm Jup}$ planets, respectively, relative to the homogenized cases. We also compare the radii with the models of \citet{Fortney2007}, which considered giant exoplanets with a compact $25\,M_{\oplus}$ core at 1 AU. Their models predict radii that are larger by $2.5$ to $3.5\%$ compared to the fuzzy-core models presented here.}
    \label{fig:final_fuzzy_core_survive}
\end{figure*}

We compare evolutionary models for giant exoplanets that incorporate updated treatments of heavy-element distributions, updated EOSs, and new atmospheric boundary conditions against the heritage Sonora Bobcat models \citep{Marley2021}. In our updated calculations, heavy elements are treated explicitly with the AQUA EOS, hydrogen–helium is modeled with the latest CD21 EOS \citep{Chabrier2021}, and the atmospheric boundary conditions are taken from the updated grids of \citet{Chen2023}. Helium rain is also included, utilizing the hydrogen–helium miscibility curve of \citet{Lorenzen2009,Lorenzen2011}, shifted by +300 K, which simultaneously reproduces the observed cooling histories of Jupiter and Saturn with fuzzy cores \citep{Sur2025a}. The initial fuzzy-core configurations in this study are constructed following the recent mass-bulk metallicity relation of \citet{Chachan2025}, with cold starting entropies chosen to preserve composition gradients throughout the evolution. We present representative models for total masses of $0.3$, $1$, $2$, and $4\,M_{\rm Jup}$, which correspond to heavy-element masses of $21$, $42$, $71$, and $128\,M_{\oplus}$, respectively \citep{Chachan2025}. All models include a compact core of $3M_\oplus$, are non-rotating, and are not tuned to reproduce any particular observable, thereby providing a controlled comparison with earlier homogeneous models. For additional context, we also include Jupiter and Saturn fits from \citet{Sur2025a}, which differ from the other cases in that they include rotation and are explicitly tuned to match both the thermal evolution and the low-order gravitational moments of the two planets. Therefore, our 1 $M_{\rm Jup}$ and 0.3 $M_{\rm Jup}$ planets presented in this work are not updated evolutionary models for Jupiter and Saturn, but rather representative exoplanet analogs at those masses that incorporate all the physical upgrades discussed herein.

Our emulated comparison Sonora Bobcat models assume a homogeneous $3.16\,Z_{\odot}$ envelope, where the influence of heavy elements is not modeled with a dedicated EOS, but is instead approximated by adopting an effective helium fraction $Y^{\prime} = Y + Z = 0.3219$ as in \citet{Marley2021}, uniformly distributed in the envelope. Neither helium rain nor the evolution of an initial fuzzy core are considered for the Sonora models. This distinction highlights a key difference: while the heritage models capture bulk metallicity in an approximate way and assume homogeneity in the envelope, our calculations account for the radial distribution of heavy elements and their impact on planetary structure and cooling. The goal of this overall exercise is to demonstrate the differences between heritage evolutionary models now used widely and more sophisticated models, however provisional.

The results, shown in 
Figure \ref{fig:final_teff_rad_masses}, reveal systematic differences between the homogeneous (dashed lines) and fuzzy core models (solid lines). At 5 Gyr, the effective temperature offsets relative to the Sonora Bobcat sequences are $18$, $1.8$, $7.4$, and $8.3$ K for the $0.3$, $1$, $2$, and $4\,M_{\rm Jup}$ planets, respectively, while the radii differ by only $\sim 1\%$. Thus, the effect of more realistic interior structures is subtle in radius, but more noticeable in cooling rates, particularly at low masses. The lower panels of Figure \ref{fig:final_teff_rad_masses} show that luminosities are nearly identical for the higher-mass planets (except at very late times), but the $0.3\,M_{\rm Jup}$ case is up to an order of magnitude brighter than its homogeneous counterpart, underscoring the sensitivity of lower-mass planets to heavy-element distribution and helium rain. The outer envelope metallicities also diverge significantly with time, reflecting the role of interior mixing and the redistribution of heavy elements. Collectively, these results demonstrate that incorporating realistic microphysics and heavy-element distributions produces modest changes at Jupiter mass and above, but can substantially alter the evolutionary trajectories of low-mass giant planets.

Next, we compare models in which the fuzzy core survives during 10 Gyr of evolution to those in which it rapidly homogenizes for planets of mass $0.3$, $1$, and $2\,M_{\rm Jup}$. All models begin with an initially extended fuzzy core encompassing roughly $30\%$ of the total planetary mass, but differ in their initial entropy profiles. Cold-start configurations, with interior entropies of $\sim 6$–$7\,k_{\rm B}$/baryon, maintain their compositional gradients and evolve with stable fuzzy cores, whereas hot-start models, with entropies of $\sim 8$–$8.5\,k_{\rm B}$/baryon, experience vigorous mixing that erases the gradient and produces nearly homogeneous interiors. The luminosity evolution for all three masses is shown in the top-left panel, while the top-right panel highlights the evolution of atmospheric metallicities, which increase substantially, by approximately $3$–$5\, Z_{\odot}$ in the homogenized models as a result of enhanced convective mixing. The middle panels of Figure \ref{fig:final_fuzzy_core_survive} illustrate this evolution through the heavy-element mass fraction and entropy distributions at 0.05 Gyr (translucent lines) and 5 Gyr (dark lines), showing the rapid homogenization in the hot-start cases. 

Quantitatively, the cold-start models with surviving fuzzy cores exhibit modest differences in their thermal and structural properties compared to the homogenized cases. At 5 Gyr, their effective temperatures differ by $-4.0$ K, $-9.1$ K, and $+4.8$ K, while the corresponding radii differ by $0.11\%$, $-0.05\%$, and $+1.54\%$ for the $0.3$, $1$, and $2\,M_{\rm Jup}$ planets, respectively. For additional context, we compare these results to the compact-core models of \citet{Fortney2007}, who considered giant planets with a $25\,M_{\oplus}$ solid core at 1 AU. Their models predict radii larger by $2.5$–$3.5\%$ relative to our fuzzy-core models, indicating that extended heavy-element distributions produce slightly more compact structures, but are also sensitive to the bulk metallicities. Overall, these results demonstrate that the survival or erosion of fuzzy cores introduces measurable, but subtle, differences in planetary cooling histories, while strongly influencing atmospheric metallicities—a potentially observable diagnostic of interior mixing and formation history \citep{Knierim2024, Knierim2025}.

\begin{figure*}
    \centering
    \includegraphics[width=0.94\linewidth]{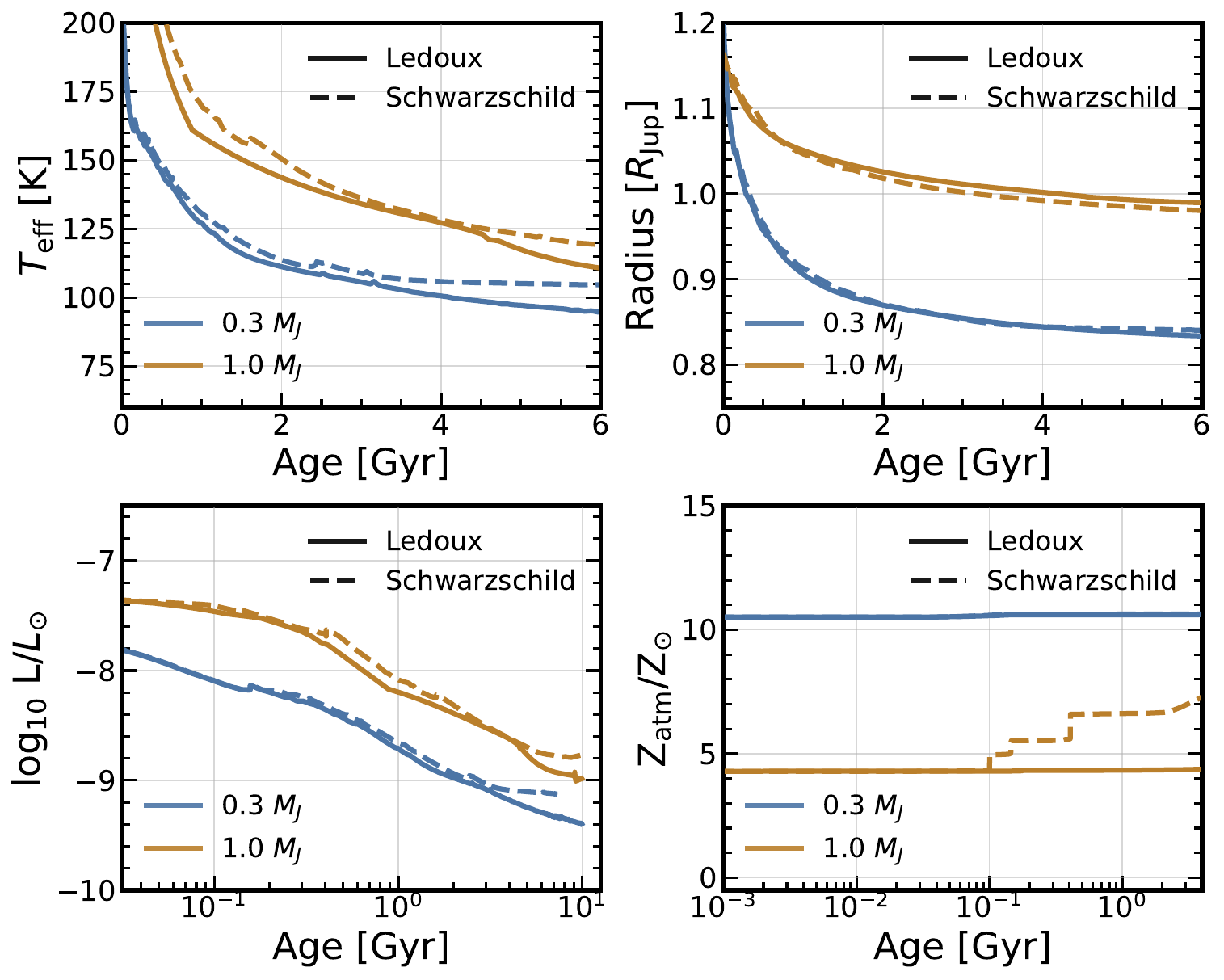}
    \caption{Comparison of evolutionary models using the Ledoux (solid lines) and Schwarzschild (dashed lines) convection criteria for $0.3\, M_{\rm Jup}$ and $1\, M_{\rm Jup}$ exoplanets, incorporating the combined updated physics. All models begin with an initial fuzzy core extending to 30\% of the normalized planetary mass and include helium rain based on the miscibility curve of L0911 shifted by +300 K. Relative to the Ledoux cases, the Schwarzschild models evolve to slightly higher effective temperatures—hotter by about $5$–$10$ K, while showing negligible differences in radius. The onset of helium rain produces a marked divergence between the two convection treatments: in Ledoux models, helium rain stabilizes the interior and suppresses convection, leading to cooler $T_{\rm eff}$ tracks, whereas in Schwarzschild models, the absence of compositional stabilization allows continued mixing and a transient increase in effective temperature after helium rain. The Schwarzschild criterion also enables more efficient redistribution of heavy elements, resulting in higher atmospheric metallicities, particularly for the $1\,M_{\rm Jup}$ planet, where the atmosphere is enriched by approximately $3\,Z_{\odot}$.}
    \label{fig:final_ldx_schw}
\end{figure*}
Finally, the choice of convection criterion also introduces systematic, but measurable, differences in planetary evolution as shown in Figure \ref{fig:final_ldx_schw}. For both the $0.3\,M_{\rm Jup}$ and $1\,M_{\rm Jup}$ cases, models using the Schwarzschild criterion evolve to slightly hotter tracks than those adopting the Ledoux criterion, with effective temperatures higher by about $5$–$10$ K at Gyr ages. The radius response is comparatively modest, shrinking by only $\sim 0.1\%$. More importantly, the treatment of convection affects the efficiency of mixing: the Schwarzschild criterion permits greater redistribution of heavy elements within the envelope, leading to a noticeable increase in atmospheric metallicity over time. This effect is especially pronounced in the $1\,M_{\rm Jup}$ case, where enhanced mixing drives surface abundances to significantly higher values than in the Ledoux models. 

The onset of helium rain further accentuates the differences between the two convection criteria. In Ledoux models, the compositional gradient produced by helium phase separation stabilizes the fluid against convection, reducing the efficiency of heat transport and leading to a drop in effective temperature. By contrast, the Schwarzschild models allow convection to persist even during helium rain, which facilitates more efficient energy release from the interior and temporarily raises $T_{\rm eff}$. As a result, the two treatments yield divergent thermal histories and surface compositions, despite their nearly identical global radii. These results highlight that, while structural differences may appear small, the choice of convection criterion plays a critical role in controlling how helium rain and internal mixing shape the observable properties of giant planets.

\textcolor{black}{Additionally, we note that the Schwarzschild criterion should not be used to transport composition across stabilizing composition gradients, as the Ledoux criterion is the correct condition. However, semi-convective or double-diffusive processes indeed operate with mixing efficiencies weaker than fully developed convection \citep{Leconte2012} and semi-convective staircases can be eroded by convective overshoot, potentially restoring fully efficient convection on shorter timescales \citep{Moll2017Double-diffusiveJupiter,Tulekeyev2024a,Garaud2018}. In \citet{Sur2025b}, we explore the use of the Schwarzschild criterion as an upper-limit case representing maximal mixing for comparison against the Ledoux scenario, thereby bracketing the range of possible evolutionary behaviors in the face of large uncertainties in semi-convective transport.}

\section{Conclusion and Discussions}\label{sec:conclusion}

In this paper, we compared old models of giant exoplanets with new models incorporating the latest physical insights, as determined from detailed modeling of Jupiter and Saturn. These new aspects
include new equations of state, helium rain, ``fuzzy" extended heavy-element cores, stably-stratified regions, and inhomogeneous heavy-element distributions in the envelope. Also included are new atmosphere boundary conditions, which factor in ammonia clouds. First, we isolated the differences between old and new models for each new feature individually. Then, we combined the new features into representative baseline models and determined the overall differences in model observables and in the interiors of all new features collectively. To make these comparisons, we created emulators of these past models using our new code \texttt{APPLE}. What we found were important variations between old and new, which should be addressed in the next generation of giant exoplanet models.  We also found that the initial thermal profiles, particularly in the interiors, can have important effects on the temporal evolution of observables, such as effective temperature and atmospheric metallicity. We focused
on the planet mass range from 0.3 to 4.0 \mjup.

Our study is meant to highlight the need for a new generation of giant exoplanet evolutionary models, but we did not in this paper provide such a new model suite and are deferring that task to later work. We also note that the potential effects of semiconvection were not explicitly addressed, though we did contrast models using the Ledoux and Schwarzschild convective criteria. The latter is in the limit of very efficient doubly-diffusive mixing, while the former mutes it completely and allows negative composition gradients to be stabilizing. We find that whether helium rain in the lower-mass giants heats
or cools their atmospheres hinges upon which limit obtains.

Our simulations have limitations.  Foremost is the ongoing need for consistent equations of state for ternary mixtures.  We have been forced to use the additive volume law. In addition, databases in the literature for individual constituents are thermodynamically consistent only to percents \citep{Tejada2024}. \textcolor{black}{Also, rotation can affect the size and evolution of the convective envelope and mixing \citep{Fuentes2023,Fuentes2024}. However, its primary importance lies in fitting gravitational moments which are lacking for giant exoplanets. It's impact on the thermodynamic structure and luminosity evolution is expected to be minor, and we therefore neglect rotation in the present calculations.} Moreover, we need a better model for hydrogen/helium miscibility and the dynamics and extent of the helium rain region. Finally, as noted, the potential role of semiconvection in thermal and material transport \citep{Leconte2012, Polman2024} is not yet well understood. Nevertheless, we have demonstrated that heritage evolutionary models of giant exoplanets since the early work of \citet{Burrows1995}, which assumed among other things, homogeneity and adiabaticity, are in need of retooling so that what emerges can play a productive role in the emerging era of precision exoplanetary science. 

\section*{acknowledgments}   
Funding for this research was provided by the Center for Matter at Atomic Pressures (CMAP), a National Science Foundation (NSF) Physics Frontier Center, under Award PHY-2020249. Any opinions, findings, conclusions, or recommendations expressed in this material are those of the author(s) and do not necessarily reflect those of the National Science Foundation. 

\appendix
\section*{The effect of old vs. new AQUA EOS}

\textcolor{black}{The original \citet{Mazevet2019} EOS had an error in its entropy term, which is used in the AQUA EOS \citep{Haldemann2020}. It was later corrected in \cite{Mazevet2021}. The impact of this error on the evolution has already been explored by \citet{Tejada2025b} in the context of Uranus (using the same code \texttt{APPLE}), where the water EOS is far more influential in its evolution than it would be for gas giants.  \citet{Tejada2025b} compared the evolution of Uranus with the AQUA EOS and an updated AQUA EOS that includes the \citet{Mazevet2021} corrections, finding that the temperatures at a given entropy of the updated EOS are higher, leading to a larger radius by about 2\%. We argue here that if such a correction is minor in the context of Uranus, which is expected to harbor most of its mass in heavy metals (such as, perhaps, water; 80–95\%), its impact is even less significant in gas giants, where the metal content comprises a much smaller fraction of the total mass ($\sim$10–12\%). For completeness, we present here two homogeneous evolutionary models with and without the \citet{Mazevet2021} corrections and show that the differences are below 0.3\% in evolution.}

\begin{figure}[H]
    \centering
    \includegraphics[width=1.0\linewidth]{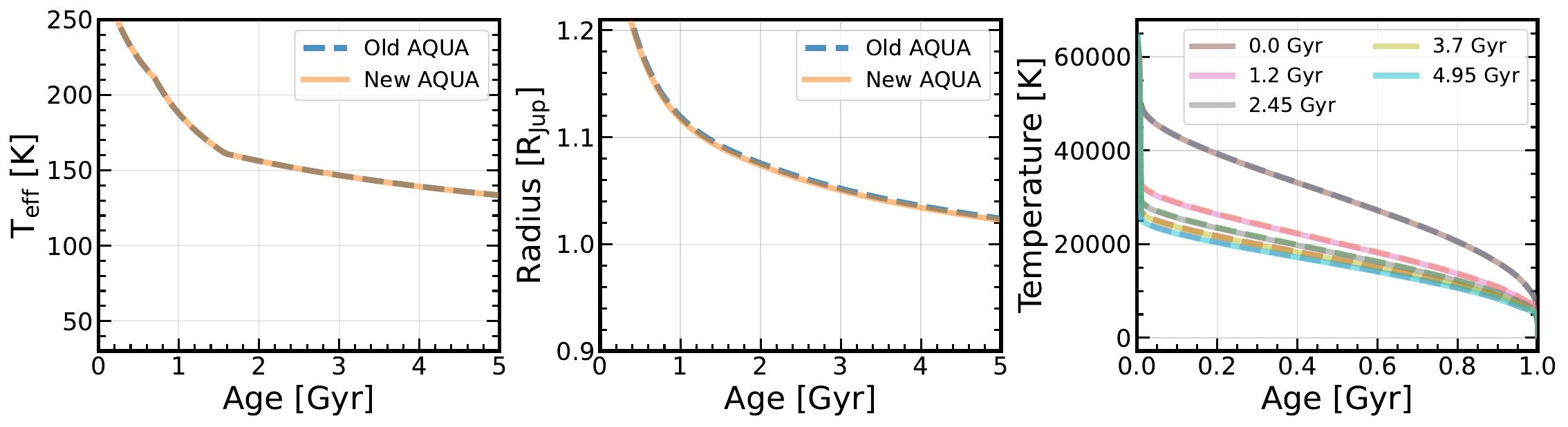}
    \caption{\textcolor{black}{Adiabatic evolutionary models for a 1 Jupiter mass planet comparing the old AQUA EOS (dashed lines) with the updated AQUA EOS (solid lines). The left panel shows that the effective temperature evolution differs by less than 0.1\%, the middle panel shows that the radius differences are below 0.3\%, and the right panel compares temperature profiles at different evolutionary times, which maximally differ by less than 1\%.}}
    \label{fig:error_aqua}
\end{figure}

\bibliography{references}{}
\bibliographystyle{aasjournal}

\end{document}